\documentclass[12pt]{article}
\usepackage[T1]{fontenc}
\usepackage{amssymb,amsthm,amsmath,mathrsfs,bm}

\begin{document}

\title{{ The matrix  Lax representation of the generalized Riemann equations and its    conservation laws}}

\maketitle

\begin{center}
  
Ziemowit Popowicz

Institute of Theoretical Physics, University of Wroc{\l}aw, pl. M. Borna 9,

50-205, Wroc{\l}aw, Poland

tel. 48-71-375-9353, fax 48-71-321-4454
ziemek@ift.uni.wroc.pl

\end{center}

\begin{abstract}
It is  shown that the generalized Riemann equation  is equivalent with the 
multicomponent generalization of the Hunter - Saxton equation. New matrix and scalar Lax representation 
is presented for this generalization. New class of the conserved densities,  which depends  explicitly on the time are obtained directly from the Lax operator. The algorithm,  which allows us to  generate a big class of the non-polynomial  conservation laws of the generalized Riemann equation  is presented. Due to this new series of conservation laws of the Hunter-Saxton equation is obtained. 
\end{abstract}

\section*{Introduction.}
The theory of the hydrodynamical type systems of the non-linear equations \cite{Whit} , integrable by the generalized
hodograph method \cite{Car} is closely related to the over-determined systems of first order
partial differential linear equations . It is achieved introducing the so called Riemann invariants in which the hydrodynamic type system is rewritten in the diagonal form as $r^i_t = \mu(r)^i r^i_x$
where $i=1,2\dots, r=(r_1,r_2...r_N)$ and no summation on the repeated indices. Thus, it  is a linear systems of first order partial differential equations with variable coefficients $\mu(r)$.

	When $N=1$ the equation  on Riemann invariant reduces to the so called Riemann equation 
$r_t=-rr_x$ which have been investigated in many papers and could be considered as the dispersionless 
limit of the Korteweg de Vries equation \cite{Ablo}. Recently the interesting generalization of the Riemann equation to the multicomponent case $(\partial /\partial t + u \partial/\partial x)^N u =0, N=1,2 \dots$ have been proposed in \cite{Pop1,Pop2,Pop3}

	When $N=2$ this generalized system is reduced to the Gurevich-Zybin system \cite{Gur1,Gur2}  or to the
 equation which describes the non-local gas dynamic \cite{Das} or to the Whitham type system \cite{Whit}. It is possible also to reduces the  $N=2$ generalized Riemann equation \cite{Pop1}  to the celebrated Hunter-Saxton equation \cite{Hunt1}, sometimes referred as the Hunter-Zheng equation \cite{Hunt2}. The Hunter-Saxton equation has been studied in almost all respects, including its complete solvability by quadratures \cite{Sak,Pav1}, construction of an infinite number of conservation laws \cite{Hunt1,Hunt2,Wang}, relationship with the Camassa-Holm  equation and the Liouville equation \cite{Pav2}, Bi-Hamiltonian formulation \cite{Hunt2,Pav2}, integrable finite-dimensional reductions \cite{Hunt2,Beal}, global solution properties \cite{Hunt3,Bres}, to mention only a few of numerous publications on this equation.
	
	For an arbitrary  $N$ the investigation of the properties of the generalized Riemann equation just 
started in \cite{Pop1,Pop2,Pop3}. It  was  indicated  that   $N=3$  generalized Riemann equation possess the matrix Lax representation, the Hamiltonian formulations and  huge number of polynomial and non-polynomial conservation laws. However this  matrix Lax representation  is a free-form,  because it contains one arbitrary function which  could be fixed but  in not a unique manner, taking into account the integrability condition on the Lax representation.

	In this paper we present the  matrix Lax representation for an arbitrary $N$ which is not a free-form.
This representation for $N=2$ could be reduced to the very well known energy-dependent second-order  Lax operator \cite{Fordy1,Aratyn,Pav1} while  for $N=3$  to the  energy depended third-order Lax operator introduced in \cite{Fordy2}. Some of the functions,  which constitute   the scalar Lax pair for $N=2,3$,  are also the non-polynomial conserved Hamiltonian functionals. Moreover from this  matrix Lax representation  it is possible to obtain the  conserved densities  which are explicitly  time depended. We present the operator, in some sense the analogue of the recursion operator,  which generates an infinite number of non-polynomial conservation laws. As the by-product of our analysis we present new  series of the non-polynomial conservation laws for  the Hunter-Saxton equation. 

	The paper is organised as follows. The first section describes the generalized Riemann equation and
 shows its connection with  the multi-component generalizations of the Hunter-Saxton equation. In the second
 section we define new matrix representation for an arbitrary $N$ extended Riemann equation. The third section
 describes the reduction of matrix Lax representation, for $N=2$ and $3$, to the scalar Lax representation. In the fourth section the
 conservation laws for $N=2,3$ generalized Riemann equation  are obtained directly from the Lax representation. The fifth section describes the
 algorithm of generations of the non-polynomial conservation laws. 

\section{Generalized Riemann Equation.}

The hydrodynamical Riemann equation 
\begin{equation}\label{riem1}
u_{t}  =   - uu_{x},
\end{equation}
have been recently generalized to the multicomponent case \cite{Pop1,Pop2,Pop3}   
as 
\begin{equation}\label{riemN}
{\cal D}_{t}^{N} u =0,\quad ~~~~~ {\cal D}_{t} = \partial / \partial t + u \partial  / \partial x,
\end{equation}
where $N=1,2,3\dots $. We can rewrite the last equation to the more comfortable form, introducing the notation $u_1=u,u_n={\cal D}^{n-1}u$ and then the $N$  generalized Riemann equation takes the following form
\begin{eqnarray} 
u_{1,t} &=&  u_2 - u_1u_{1,x}, \\ \nonumber 
u_{2,t} &=&  u_3 - u_1u_{2,x}, \\ \nonumber 
........&&......................... \\ \nonumber
u_{N-1,t} &=& u_{N} - u_1u_{N-1,x},  \\ \nonumber 
u_{N,t} &=& - u_1u_{N,x}.
\end{eqnarray}

The  multicomponent generalization of the Hunter-Saxton equation could  be  obtained from the last formula using the transformation $ u_{2,x} = (u_{1,x}^2 + w_2^2)/2, u_{k,x} = w_2^{2} w_{k}$ for $k=3,4,\dots$.
\begin{eqnarray}
u_{1,t,x} &=& -\frac{u_{1,x}^2}{2} - u_{1}u_{1,xx} +\frac{ w_2^2}{2}, \\ \nonumber 
w_{2,t} &=& -(u_1w_2)_x + w_3 w_2, \\ \nonumber
w_{3,t} &=& u_{1,x}w_3 - u_1w_{3,x} -2w_3^2 + w_4, \\ \nonumber 
........&&........... \\ \nonumber 
w_{N-1,t} &=& u_{1,x}w_N - u_1w_{N-1,x} - 2w_3w_{N-1} + w_{N},  \\ \nonumber 
w_{N,t} &=& u_{1,x}w_{N} - u_{1}w_{N,x} -2w_3w_{N}.
\end{eqnarray}
When all $w_i$ vanishes then Esq. (4) reduces to th Hunter-Saxton equation while for $w_i=0, i=3,4,\dots$ our
equations reduces to the two-component Hunter-Saxton equation

\section{Matrix Lax  representation.}
The generalized Riemann  equation for $N > 1$ could  be obtained  from  the compatibility condition 
for the matrix Lax representation 
\begin{equation}
 \Psi_x = A \Psi, \quad ~~~~~ \Psi_t = -u_1 A\Psi - \lambda E\Psi,
\end{equation}
where $\Psi=(\psi_1,\psi_2,\dots \psi_n)^{T}$, $A,E$ are  $ N\times N $ matrices such that $E_{k,k-1}=1$ 
for $k=1,2,\dots N-1$  and 
\begin{eqnarray} 
 A= \left( \begin{array}{cccccc}
           \lambda^2u_{N-1,x} & \lambda u_{N,x} & 0 & \dots  &  0 \\
0 & \lambda^2u_{N-1,x} & 2\lambda u_{N,x} &  \dots & 0 \\
  & :  & : & : &  & \\
 \dots & \dots  & \lambda^2u_{N-1,x} & k\lambda u_{N,x} & \dots  \\
 &  : & : & : & & \\
-N\lambda^{N+1} & -N\lambda^{N}u_{1,x}   & \dots & - N\lambda^3u_{N-2,x}  & -(N-1)\lambda^2u_{N-1,x}  \\
           \end{array}\right ) 
\end{eqnarray}

Explicitly for $N=2,3$ we obtain 
\begin{equation} \label{mat2}
\left( \begin{array}{cc}\psi_1 \\ \psi_2 \end{array} \right )_x
=\left( \begin{array}{cc}
             \lambda^2 u_{1,x}& \lambda u_{2,x}  \\
	-2\lambda^3 & -\lambda^2 u_{1,x} 
            \end{array}\right )
\left( \begin{array}{cc}\psi_1 \\ \psi_2 \end{array} \right ),
\end{equation}
\begin{equation}
\left( \begin{array}{cc}\psi_1 \\ \psi_2 \end{array} \right )_t
 = \left(
\begin{array}{cc}
  -\lambda^2 u_1u_{1,x}  &  -\lambda u_1 u_{2,x} \\
\lambda (2\lambda^2 u_1 -1) & \lambda^2 u_{1} u_{1,x} 
\end{array} \right)
\left( \begin{array}{cc}\psi_1 \\ \psi_2 \end{array} \right ).
\end{equation}

\begin{equation} \label{mat3}
\left( \begin{array}{ccc}\psi_1 \\ \psi_2 \\ \psi_3\end{array} \right )_x
=\left( \begin{array}{ccc}
             \lambda^2 u_{2,x}  & \lambda u_{3,x} & 0 \\
	0 &  \lambda^2 u_{2,x}  & 2\lambda u_{3,x} \\
 -3\lambda^4 & -3\lambda^3 u_{1,x} & -2\lambda^2 u_{2,x} 
            \end{array}\right )
\left( \begin{array}{ccc}\psi_1 \\ \psi_2 \\ \psi_3 \end{array} \right ),
\end{equation}
\begin{equation}
\left( \begin{array}{ccc}\psi_1 \\ \psi_2 \\ \psi_3\end{array} \right )_t
=\left( \begin{array}{ccc}
             -\lambda^2u_1 u_{2,x}  & -\lambda u_1 u_{3,x} & 0 \\
	-\lambda  &  - \lambda^2 u_1 u_{2,x}  & - 2u_1 \lambda u_{3,x}  \\
 3\lambda^4u_1 & 3\lambda^3 u_1u_{1,x}- \lambda  & 2\lambda^2 u_1 u_{2,x}
            \end{array}\right )
\left( \begin{array}{ccc}\psi_1 \\ \psi_2 \\ \psi_3 \end{array} \right ).
\end{equation}

\section{Scalar Lax representation for N=2,3.}

We consider the case $N=2$ and $N=3$ separately.

\subsection{ N=2.}

	We can obtain two different  scalar Lax representation for $N=2$ generalized Riemann equation .  

For  the first choice   computing $\Psi_1$ from the second equation in the formula \ref{mat2} we obtain 
\begin{eqnarray}
 \Psi_{2,xx} &=&  \lambda^2 (-2 \lambda^2 u_{2,x} - u_{1,xx} + \lambda^2 u_{1,x}^2) \Psi,  \\ \nonumber
 \Psi_{2,t} &=& (\frac{1}{2\lambda^2} - u_1) \Psi_{2,x} + \frac{1}{2}u_{1,x}\Psi_2.
\end{eqnarray}
It is exactly the Lax operator considered in  \cite{Aratyn,Pav1}.

For the second case computing $\psi_2$ from the first equation in \ref{mat2} and using the transformation 
$\psi_1 \Rightarrow  \sqrt{u_{2}} ~~\varPhi$ we obtain 
\begin{eqnarray} 
 \varPhi_{xx} &=& ( \lambda^4  Z_2 +  \lambda^2 Z_1 + Z_0)\varPhi, \\ \nonumber 
 \varPhi_t &=& ( u_{1,x}\varPhi -2u_1\varPhi_x)/2.
\end{eqnarray}
where 
\begin{equation}
 Z_2 = -2u_{2,x}  + u_{1,x}^2,\quad 
 Z_1 = u_{2,x}(\frac{u_{1,x}}{u_{2,x}})_x, \quad
 Z_0 = -\frac{1}{4} \frac{2u_{2,xxx}u_{2,x} - 3u_{2,xx}^2}{u_{2,x}^2}.
\end{equation}

The integrability condition $\varPsi_{xx,t} = \varPsi_{t,xx}$ gives us 
\begin{eqnarray}\label{equ1}
 Z_{i,t}&=& -(\partial Z_i + Z_i\partial ) u_1, ~~~~~~~for~~ i=1,2 ,\\ \nonumber 
 Z_{0,t}&=& (\frac{1}{4} \partial^3  -2\partial Z_0 -2 Z_0\partial ))  u_1.
\end{eqnarray}
and it  is identically satisfied if we use the definition of $Z_2,Z_1,Z_0$. 

\subsection{\textbf{$N=3$.}}

Computing  $\psi_2$ and $\psi_3$ from the first and second equation in \ref{mat3} respectively   and using the transformation 
$\psi_1 \Rightarrow u_{3,x}\Omega $ we obtain 
\begin{eqnarray}
 \Omega_{xxx} &=& \sum_{k=0}^2 \lambda^{2k} S_k\Omega_x + (\sum_{k=0}^3\lambda^{2k}P_k +\frac{1}{2}\sum_{k=0}^2
\lambda^{2k}S_{k,x})\Omega ,\\ \nonumber 
\Omega_t &=&  u_{1,x}\Omega -u_1\Omega_x.
\end{eqnarray}
where 
\begin{eqnarray}
 P_0 &=& 0, ~~~~~~~~~~~\quad ~~~~~~~~~~~~~ P_3 = 2(3u_{1,x}u_{3,x}u_{2,x} - 3u_{3,x}^2 - u_{2,x}^3), \\ \nonumber
 P_2 &=& 3\frac{u_{3,xx}u_{2,x}^2}{u_{3,x}} -  3(u_{1,x}u_{3,xx} - u_{3,x}u_{1,xx} + u_{2,x}u_{2,xx}), \\ \nonumber
 P_1 &=& -\frac{1}{2}( \frac{3u_{3,xx}^2 u_{2,x}}{u_{3,x}^2} - \frac{1}{u_{3,x}}( u_{3,xxx}u_{2,x}   + 3u_{2,xx}u_{3,xx})  + u_{2,xxx}), \\ \nonumber
 S_2 &=& 3(u_{2,x}^2 - 2u_{1,x}u_{3,x}), ~~~~~~~~
 S_1 = 3\frac{1}{u_{3,x}}  (u_{2,xx}u_{3,x}  - u_{3,xx}u_{2,x}), \\ \nonumber
 S_0 &=& \frac{1}{u_{3,x}^2} (3u_{3,xx}^2 - 2u_{3,xxx}u_{3,x}). 
\end{eqnarray}
This is the energy-dependent third-order Lax operator considered in \cite{Fordy2}.

From the integrability condition  we obtained the following  equations of motion on  the functions $S_i,P_i$  
\begin{eqnarray}\label{equ2}
P_{i,t} &=& -(\partial P_i  + 2 P_i \partial) u_1, \quad for ~~~i=1,2,3, \\ \nonumber
S_{i,t} &=& -(\partial S_i  +  S_i \partial) u_1, \quad  for ~~~ i=2,1,  \\ \nonumber
S_{0,t} &=& (\partial^3 - \partial S_0  -  S_0 \partial) u_1.
\end{eqnarray}

For the second choice where we eliminate   $\psi_3$ and $\psi_1$ from the second  and third  equation in  \ref{mat3} respectively   and using the transformation $\psi_2 \Rightarrow u_{3,x}^{\frac{2}{3}}\Omega $ we obtain 
\begin{eqnarray}
 \Omega_{xxx} &=& \sum_{k=0}^2 \lambda^{2k} \hat S_k \Omega_x+ \sum_{k=0}^3 \lambda^{2k}\hat P_k \Omega, \\ \nonumber 
\Omega_t &=&  \lambda^{-4} W_0\Omega_{xx} + \sum_{k=0}^2 \lambda^{-2(2-k)} W_{k+1}\Omega_x +  
\sum_{k=0}^2 \lambda^{-2(2-k)}W_{k+4}\Omega.
\end{eqnarray}

where 
\begin{eqnarray} 
 \hat S_0 &=& \frac{1}{u_{3,x}} (4u_{3,xx}^2 - 3u_{3,xxx}u_{3,x}), \\ \nonumber 
 \hat S_1 &=& \frac{-1}{u_{3,x}}  u_{3,xx}u_{2,x} , ~~~~~~~~~~~~~~~~~~~~~~~~~~
 \hat S_2 =  3u_{2,x}^2 - 6u_{3,x}u_{1,x} 
\end{eqnarray}

\begin{eqnarray}
 \hat P_0 &=& \frac{1}{ 3u_{3,x}^2} (72u_{3,xxx}u_{3,xx}u_{3,x} - 18u_{3,4x}u_{3,x}^2 - 
 56u_{3,xx}^3), \\ \nonumber 
\hat P_1 &=& \frac{1}{  3u_{3,x}^2} (4u_{2,x}u_{3,xx}^2 + 3u_{3,x}^2u_{2,xxx} -3u_{3,x}(u_{3,xxx}u_{2,x} + 2u_{3,xx}u_{2,xx})) , \\ \nonumber 
\hat P_2 &=& \frac{1}{3 u_{3,x}}  (u_{2,x}(- 3u_{3,xxx}u_{3,x} + 4u_{3,xx}^2) - 
u_{2,xx}(6u_{3,xx}u_{3,x} + 3u_{3,x}^2)), \\ \nonumber
\hat P_3 &=& (u_{3,xx}(-4u_{3,x}u_{1,x} + u_{2,x}^2)-3u_{3,x}(2u_{3,x}u_{1,xx} - u_{2,xx}u_{2,x})) .
\end{eqnarray} 

\begin{eqnarray} 
W_0 & =& \frac{1}{6u_{3,x}},~~~~~~~~~ W_1 =\frac{u_{3,xx}}{18u_{3,x}^2}, ~~~~~~~~W_2 = u_{2,x}W_0,~~~~~~~~~ W_3=-u_1, \\ \nonumber 
W_4 &=& \frac{1}{27u_{3,x}^3} (3u_{3,xxx}u_{3,x} - 4u_{3,xx}^2 ),~~~~~~~
W_5 = \frac{1}{18u_{3,x}^2}  (5u_{3,xx}u_{2,x} - 3u_{3,x}u_{2,xx}), ~~~~~~\\ \nonumber 
W_6 &=& \frac{1}{3u_{3,x}} (5u_{3,x}u_{1,x} - u_{2,x}^2).
\end{eqnarray}

	In that way  Esq. 18 could be  considered as the  energy-dependent third-order Lax operator also.

\section{The conservation laws obtained from Lax representation.}

The knowledge of scalar Lax representation allows us to compute the conservation laws for the 
model. 

For the $N=2$ generalized Riemann equation let us  rewrite the first Lax pair in terms of two Riccati  equations introducing the function $\Gamma = \Psi_{2,x}/\Psi_2$
\begin{eqnarray}
 \Gamma_x &=& -\Gamma^2 - \lambda^2 u_{1,xx} - \lambda^4 (2u_{2,x} - u_{1,x}^2), \\ \nonumber
 \Gamma_t &=& (u_1 - \frac{1}{2\lambda^2}) \Gamma^2 - u_{1,x}\Gamma +
 \lambda^2 (\frac{1}{2}u_{1,x}^2 + u_{1,xx}u_1 - u_{2,x}) + 
 \lambda^4 u_1(2u_{2,x} - u_{1,x}^2).
\end{eqnarray}
Expanding  $\Gamma$ in powers of $\lambda$ as
\begin{equation}
\Gamma = -\lambda^2 u_{1,x} -2\lambda^4 u_2 +\sum_{k=0}^{\infty} \lambda^{2k+6} \varGamma_{2k}.
\end{equation}
we obtain for example the first two  integrable equations 
\begin{eqnarray} 
 \varGamma_{0,t} &=&  4u_{1,x}u_2u_1 - 2u_2^2 ,\quad ~~~~~~~~~~
 \varGamma_{2,t} =  - 2(u_1u_{1,x} - u_2)  \varGamma_0 + 4u_1u_2^2, \\ \nonumber
\varGamma_{0,x} &=& -4u_2u_{1,x}, \hspace{3cm} 
\varGamma_{2,x} = 2u_{1,x}\varGamma_0 - 4u_2^2.
\end{eqnarray}
From this follows that all $\varGamma_{n,x}$  are conserved. Indeed
\begin{eqnarray} 
 H_0 &=&  \int ~ dx ~ \varGamma_{0,x}  =-4\int ~dx ~ u_{1,x}u_2, \\ \nonumber
 H_2 &=& - 4\int ~dx ~ 2u_{1,x} \partial^{-1} u_{1,x}u_2 +u_2^2 = 4\int ~dx ~u_1^2u_{2,x} - u_2^2.
\end{eqnarray}
These conserved Hamiltonian functionals  coincides with those obtained in \cite{Das}.

Quite different series of conserved  densities we obtain using the matrix Lax representation Esq. 9 and 10 in which we redefine the functions $\Psi_1$ and $\Psi_2$ as 
\begin{equation} 
 \Psi_1= e^{(g+\lambda^2 u_1)},\quad ~~~~~~ \Psi_2=\varUpsilon e^{(g+\lambda^2 u_1)}.
\end{equation}
Then our Lax representation gives us 
\begin{eqnarray}
 g_x &=& \lambda u_{2,x}\varUpsilon  , \quad ~~~~~~~~~~~~~~~~~~~~~~~~~~~~~ g_t=-u_1g_x - \lambda^2 u_{2}, \\ \nonumber 
\varUpsilon_{x} &=& -\lambda u_{2,x}\varUpsilon^2 - 2\lambda^2u_{1,x}\varUpsilon - 2\lambda^3,  \quad ~
\varUpsilon_t = -u_1\varUpsilon_x - \lambda.
\end{eqnarray}
The integrability conditions $g_{x,t}=g_{t,x}, \varUpsilon_{x,t} = \varUpsilon_{t,x}$ lead  us the $N=2$
generailzed Riemann  equation. These two equations are in the conservative form and thus $g_x$ and $\varUpsilon_x$ are conserved Hamiltonian functionals. The explicit form of these functionals   could be obtained  expanding the 
function $\varUpsilon$ as 
\begin{equation}
 \varUpsilon =\frac{1}{u_2^2}\sum_{k=0}^{\infty} \lambda^k\varrho_k,
\end{equation}
where for example 
\begin{eqnarray}
 \varrho_0&=&1, ~~~~~~~~~\varrho_1=u_2 - t, ~~~~~~~~~~~~\varrho_2=2u_2t +u_2^2 - 2u_1, \\ \nonumber
\varrho_3&=&-t^2u_2 +t(2u_1-3u_2^2) -2x +2u_1u_2 - u_2^3 + 2\partial^{-1}u_{2,x}u_1.
\end{eqnarray}

The  first nontrivial  conserved Hamiltonian functionals are 
\begin{eqnarray}
 H_1 &=& \int dx~u_{1,x} u_2^2, ~~~~~~  H_2 = \int dx~ u_{1,x}u_2^3,  	\\ \nonumber
H_3 &=& \int dx~ (u_{2,x}u_1^3 - 3u_2^2u_1  -6u_{2,x}u_1\partial^{-1}u_2).  \\ \nonumber 
\end{eqnarray}
and  explicit time dependend 
\begin{eqnarray}
 H_4 &=& \int dx~ (u_2 -u_{1,x}u_2 t), ~~~~~~H_5 = \int dx~(2u_1 - 2u_2t + u_{1,x}u_2t^2)  \\ \nonumber 
 H_6 &=& \int dx~ (u_2^2 + 2u_{2,x}u_2u_1t), ~~~~H_7 = \int dx~(u_{2,x}u_1^2 + u_{1,x}u_2^2t), \\ \nonumber
H_8 &=& \int dx~ (-2u_1^2 - 4\partial^{-1}u_2  +4t\partial^{-1}u_{2}u_{1,x}  - t^2(u_{2,x}u_1^2  + u_2^2)),
\end{eqnarray}
while $x$ and  time dependend 
\begin{eqnarray}
H_9 &=& \int dx~ t(6u_2^2u_1 + 12u_{2,x}u_1\partial^{-1}u_2 - 2u_{2,x}u_1^3) + 
6x(u_{2,x}u_1^2 + u_2^2), \\ \nonumber 
&& \hspace{2cm} 12u_1\partial^{-1}u_{2,x}u_1 - 6u_2u_1^2.
\end{eqnarray}

Now let us consider once more the equation 11  in which we redefine the functions $\Psi_i , i=1,2,3$ as 
\begin{equation}
  \Psi_1 = e^{(g+\lambda^2u_2)}, \quad ~~~ \Psi_2 = \Upsilon e^{(g+\lambda^2u_2)} ,\quad ~~~ \Psi_3 = 
\Xi e^{(g+\lambda^2 u_2)}.
\end{equation}

and as the result we obtain,  after eliminations of $\Xi$ 
\begin{eqnarray}
 g_x &=& \lambda u_{3,x} \Upsilon , \quad ~~~~~~~~~~~~~~~~~ g_t = -u_1g_x -\lambda^2 u_3, \\ \nonumber 
\Upsilon_{xx} &=&  \frac{u_{2,xx}}{u_{2,x}} \Upsilon_x - 3\lambda u_{3,x}\Upsilon \Upsilon_x - 
3\lambda^2 u_{2,x}\Upsilon_x -\lambda^2 u_{3,x}^2 \Upsilon^{3}, \\ \nonumber 
&&  -3\lambda^3u_{3,x}u_{2,x}\Upsilon^2 - 6\lambda^4u_{3,x}u_{1,x}\Upsilon - 6\lambda^5 u_{3,x} \\ \nonumber 
\Upsilon_t &=& -u_1\Upsilon_x -\lambda.
 \end{eqnarray}
The integrability condition $g_{x,t}=g_{t,x}$ and $\Upsilon_{xx,t}=\Upsilon_{t,xx}$ gives us the $N=3$ generalized Riemann equation. 
From this representation follows that the $g_x$ generates  conserved Hamiltonian functionals.
Indeed if we expand $\Upsilon$ in the power series in $\lambda$ as
\begin{equation}
 \Upsilon = \sum_{k=0}^{\infty} \lambda^k\phi_k,
\end{equation}
where for example 
\begin{eqnarray} 
 \phi_0 &=& u_3, ~~~~~~~\phi_1=-\frac{1}{2}u_3^3 - t,~~~~~~ \phi_2 = \frac{3}{2}tu_3^2 + \frac{1}{4} u_3^5 -3\partial^{-1}u_{3,x}u_2,  \\ \nonumber 
\phi_3 &=& -\frac{5}{4}tu_3^4 -\frac{1}{8} u_3^7 - \frac{3}{2} u_3^3u_2 + \frac{9}{2}u_3^2 \partial^{-1}u_{3,x}u_2 + \frac{3}{2}u_3\partial^{-1}u_{2,x}u_3^2.
\end{eqnarray}

then the coefficients standing in the same power of $\lambda$ in $g_x$  are conserved Hamiltonian functionals as for example
\begin{eqnarray}
H_1 &=& \int dx ~ u_{2,x}u_3^2 ,~~~~~~~~~~~~~ H_2= \int dx~ u_3^2(2u_{1,x}u_3 -  3u_{2,x}u_2),\\ \nonumber  H_3 &=& \int dx~ u_3^4(4u_{1,x}u_3 - 5u_2u_{2,x}),\\ \nonumber 
H_4 &=& \int dx~ 5u_3^4 + 12u_{2,x}u_3^2\partial^{-1}u_{3,x}u_1 - 12u_{1,x}u_3^2\partial^{-1}u_{3,x}u_2, \\ \nonumber
H_{5} &=& \int dx~ 10u_{2,x}u_3^3u_1 + 5u_{3,x}u_3u_2^3 +6u_3^2(u_{1,x}\partial^{-1}u_{3,x}u_2 - 
u_{2,x}\partial^{-1}u_{3,x}u_1). \\ \nonumber 
\end{eqnarray}
and  explicit time dependend
\begin{eqnarray} 
H_6 &=& \int dx~ u_3^2(-u_{2,x}t^2 + 2u_{1,x}t - 2), ~~~
H_7 = \int dx~ u_3^6(6u_{3,x}u_2t + u_{2,x}u_2) \\ \nonumber
H_8 &=& \int dx~ 10 t^2u_{3,x}u_3^3u_2 + 5tu_{1,x}u_3^4 + 12u_3^2(u_{2,x}\partial^{-1}u_{3,x}u_1 - 
u_{1,x}\partial^{-1}u_{3,x}u_2), \\ \nonumber 
H_{9} &=& \int dx~2t^2 u_{3,x}u_3^3u_2 - 2tu_{3,x}u_3^2u_2^2 - u_{2,x}u_3^2u_2. \\ \nonumber
\end{eqnarray}

\section{Non-polynomial conservation laws.} 
Notice that the equation on $Z_{1,t}$ in \eqref{equ1} can be rewritten as 
\begin{equation} 
 ({\sqrt{Z_1}})_t = -(u_1{\sqrt{Z_1}})_x.
\end{equation}
Similarly the equations on $P_{i,t}$ and on $S_{2,t},S_{1,t}$ can be rewritten as 
\begin{eqnarray} 
 (P_i^{1/3})_t &=& -(u_1P_i^{1/3})_x  \quad i=1,2,3, \\ \nonumber
 ({\sqrt{S_i}})_t &=& -(u_1{\sqrt{S_i}})_x \quad i=2,1. 
\end{eqnarray}
Therefore  $\int dx \sqrt{Z_1}$ and $\int dx \sqrt{S_i}, \int dx P_i^{1/3} $ are conserved Hamiltonian functionals for the $N=2,3$ generalized Riemann equation respectively.

Moreover notice that the following functions  
\begin{equation}
 H_{n,m} =\int dx \textbf{ \begin{large}(\end{large} } \frac{(k_1P_1 + k_2P_2+k_3P_3)^m}{(k_4S_1+k_5S_2)^n}{\textbf{ \begin{large})\end{large}} }^{\begin{Large}                   \frac{ 1}{3m-2n}  \end{Large}}. 
\end{equation}
where $n,m, k_i, i=1\dots 5$ are arbitrary constants such that $k_4=k_5 \neq 0$ and $k_1=k_2=k_3 \neq 0$ are
 the conservation laws 
for the $N=3$ generalized Riemann equation. Indeed we can easily verify it, using \eqref{equ2} showing,  that 
\begin{equation} 
 H_{n,m,t} = - (u_1H_{n,m})_x.
\end{equation}

We can construct the infinite number of the non-polynomial conservation laws   
using the analog of recursion operator as follows.

In the papers \cite{Pop1,Pop2} we proved that if some function $H$,  which depends on $u_n$ and its derivatives 
satisfy 
\begin{equation}
 H_t = -ku_{1,x}H - u_1H_x.
\end{equation}
then $\int~ dx~ H^{1/k}$ is a conserved Hamiltonian functionals for the generalized Riemann equation.

Moreover we have three  additional  lemmas 

Lemma1: If f and g satisfy
\begin{equation} 
 	f_t = - \mu u_{1,x}f - u_1f_x,\quad g_t=-\nu u_{1,x}g - u_1g_x.
\end{equation}
and $n,m$ are an arbitrary numbers such that  $n\mu \pm m\nu \neq 0$ then

 $H=\int~(f^ng^m)^{1/(n\mu \pm m\nu)}~dx $ is conserved
\vspace{0.5cm} 

Lemma2:  If $f$ and $g$ satisfy 
\begin{equation} 
 	f_t = - \mu u_{1,x}f - u_1f_x,\quad g_t=-\mu u_{1,x}g - u_1g_x, 
\end{equation}
then  $ H=\int~(f\pm m)^{1/\mu}~dx $ is conserved

\vspace{0.5cm} 

Lemma3: If $f$ satisfy 
\begin{equation} 
 f_t=-\mu u_{1,x}f - u_1f_x
\end{equation}
then $g=u_{N,x}^mf$  satisfy $ g_t =-(\mu + m)u_{1,x}g - u_1g_{x}$ and  
$H=\int~g^{\frac{1}{\mu + m}}~dx$ is conserved. 
\vspace{0.3cm} 

The proofs of these lemmas are elementary. 

	Let us define the following operator 
\begin{equation}
R_k = \frac{1}{u_{N,x}} \partial - k \frac{u_{N,xx}}{u_{N,x}^2} = u_{N,x}^{k-1}\partial \frac{1}{u_{N,x}^k}, 
\end{equation}
where $k$  is an arbitrary number but $k\neq 0$. 

We have the  following 

Theorem : If some function $H$ which depends on $u_n$ and its derivatives 
satisfy 
\begin{equation}
 H_t = -ku_{1,x}H - u_1H_x
\end{equation}
then $H_m = R_k^m H$ where $m=0,1,2 \dots$ generates the infinite number of non-polynomial dispersive and dispersionless conserved Hamiltonian functionals $\int ~ dx~H_m^{1/k}$.

Proof: We carry it by induction. For $m=0$ it is valid from the assumption because $H_0=H$. 
Now we show that 
\begin{equation}
 (H_{m+1})_t = -ku_{1,x}H_{m+1} - u_1H_{m+1,x}. 
\end{equation}
We have 
\begin{equation}
 (H_{m+1})_t = (R_kH_m)_t =R_{k,t}H_m - kR_ku_{1,x}H_m -R_ku_1H_{m,x} = -ku_{1,x}R_kH_m - u_1(R_kH_m)_x, 
\end{equation}
what can be rewritten as 
\begin{eqnarray}
 R_{k,t} &=& (\frac{1}{u_{N,x}})_{t} \partial + k(\frac{1}{u_{N,x}})_{x,t} =  [R_k,k u_{1,x}+ u_1\partial] = \\ \nonumber 
&& k(\frac{u_{1,x}}{u_{N,x}} - u_1(\frac{1}{u_{N,x}})_{x})_{x} + 
(\frac{u_{1,x}}{u_{N,x}} - u_1(\frac{1}{u_{N,x}})_x) \partial .
\end{eqnarray}
and it is identically satisfied if we compute $R_{k,t}$ using the equation of motion .

Notice that for $k=1$ we obtain trivial result because then $RH$ is a total derivative. 

	For $N=2$ we can use the functions $Z_1,Z_2$ defined in Esq.15 while for $N=3$  the functions
 $P_i,i=0,1,2,3, S_i,i=1,2$ defined in Esq. 19 respectively as the seeds solutions  in order to generates by
   $R$ operator an infinite number  of  conserved Hamiltonian functionals.

	Finally let us  discuss the problem of the existence of conservation laws  for generalized Riemann
 equation,  which are in  the form 
\begin{equation} 
 H_t = -(u_1H)_x. 
\end{equation}
Such  conservation  laws  could be easily obtained using the Lemma1. Indeed 
\begin{equation}
\int dx  H_{k,n,m}=\int dx \frac{R_k^{n}G_k}{R_{k-1}^mF_{k-1}}
\end{equation}
 is conserved Hamiltonian functionals ,  where $G_k,F_{k-1}$ are such that 
\begin{equation}
 G_{k,t} =-ku_{1,x}G_k - u_1G_{k,x}, \quad F_{k-1,t}=-(k-1)u_{1,x}F_{k-1} - u_1F_{k-1,x}
\end{equation}
Example: only for the Hunter - Saxton equation for which $R_k = u_{1,x}^{2k-2}\partial \frac{1}{u_{1,x}^{2k}}$
\begin{eqnarray}
 G_3 &=& u_{1,xx}u_{1,x}^{2},\quad F_2=u_{1,xx}, \\ \nonumber 
H_{3,0,1} &=& \frac{u_{1,xx}u_{1,x}^5}{u_{1,xxx}u_{1,x} - 4u_{1,xx}^2}, \\ \nonumber 
H_{3,2,0} &=& \frac{u_{1,4x}}{u_{1,xx}u_{1,x}} - 14\frac{u_{1,xxx}}{u_{1,x}^3} +28\frac{u_{1,xx}^2}{u_{1,x}^4},
\\ \nonumber 
H_{3,2,1} &=& \frac{u_{1,4x}u_{1,x}^2 - 14u_{1,xxx}u_{1,xx}u_{1,x} +28u_{1,x}^3} {
u_{1,xxx}u_{1,x}^2 - 4u_{1,xx}^2u_{1,x}}.
\end{eqnarray}

\begin{eqnarray}
 G_4 &=& u_{1,xx}^{\frac{3}{2}}u_{1,x}^{2},\quad F_3=u_{1,xx}^{\frac{3}{2}}, \\ \nonumber 
H_{4,2,0} &=& \frac{3u_{1,4x}}{2u_{1,xx}u_{1,x}^2} + \frac{3u_{1,xxx}^2}{4u_{1,xx}^2u_{1,x}^2} - 27\frac{u_{1,xxx}}{u_{1,x}^3}+54\frac{u_{1,xx}^2}{u_{1,x}^4}.
\end{eqnarray}

\end{document}